Gerard Keogh

**The Statistical and Econometric Analysis of Asylum Application Trends and their relationship to GDP in the EEA**

Gerard Keogh[1].

*Statistics Department, Trinity College Dublin.*

Abstract: The sharp decline in Ireland's economic performance in recent years has coincided with a recent fall in asylum applications. Simultaneously countries such as Switzerland are seeing increases in asylum numbers with evidence for greater numbers of Nigerian applicants, a group that have for some time been the largest nationality group applying in Ireland. A possible reason for this shift in asylum seeker preference is the general economic conditions here versus those in other European countries. In this paper we investigate whether this belief holds water. We model asylum applications as a function of GDP using a time varying parameter multiplicative growth model. Our results show there is an economic basis for asylum seeker preferences. We further show there is no regional basis for asylum seekers' expectation of a more favourable claim in the 'developed box' in central Europe as compared to countries on the so-called 'periphery'.

---

[1] Email: GMKeogh@justice.ie

## 1. Introduction

Modern migration tends to involve considerable temporary movements for family purposes, work or study and therefore the impact of migration on the home and host countries can differ to that observed in permanent migration, see Barret & Goggin (2010). For example, temporary migrants may repatriate a lot more of their income than long term permanent migrants. Nevertheless, the link between permanent or temporary migration and incomes or economic well being is well established and underpinned by the reality that migration occurs if the net present value of earnings in the host country, exceeds that of the migrants home country, see Sjaastad (1962).

In contrast the reasons for asylum movements are less well understood. Research on asylum tends to be advocacy based and focused on persons fleeing persecution with the overall objective of instigating change, Hughes & Quinn (2004). The evidence for this is clear from numerous articles in the past few years in, for example, the Journal of Refugee Studies. As a consequence, a balanced view of the true reasons behind asylum migration remains unclear. Clearly then, some persons are refugees but others who claim to be are not. These may well be asylum applicants who are in fact economic migrants and the reality underpinning their movement is simply the expectation of better incomes. In this paper we examine this claim. An economic motivation may also partly explain why some asylum seekers look to 'cheat the system', see Kibreab (2004).

Asylum seekers throughout the EEA make an application for international protection that is considered in the context of the 1951 UN Convention relating to the Status of Refugees and its 1967 Protocol. Individual states make their determination and grant or refuse refugee status. While the UNHCR monitors the determination procedure in each country, rates of success reported to Eurostat vary across the EEA. Among the main reasons for this are the procedural basis for decision and methods of reporting are not identical - some countries for example have more elaborate procedures than others. Detailed accounts of individual country procedures are set out in the Asylum Procedures Manual (IGC 2009).

### Figure 1

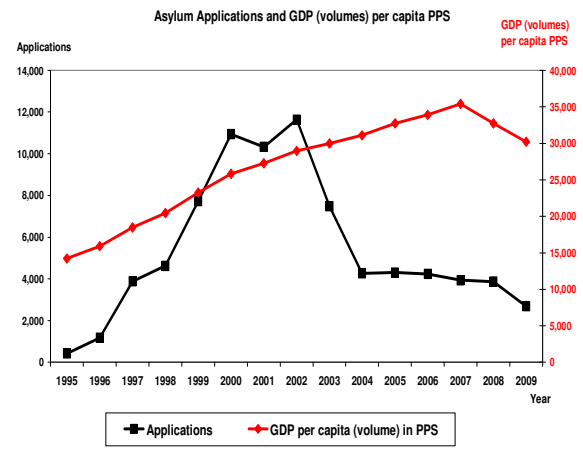

Of course, asylum applications are determined by many factors. In Ireland's case, constitutional and legislative changes such as the amendment on eligibility for citizenship, the introduction of carrier liability, the habitual residency condition and streamlining of processing procedures, permanently reduce the overall level. In contrast, economic conditions persistent and therefore may be a key factor determining application numbers over time.

The empirical evidence for a link between economic well being and asylum applications is perceptible in Ireland. In the early nineties asylum application numbers in Ireland were negligible. Through the second half of that decade numbers rose steadily to reach a peak of over 11,635 in 2002 (see Figure 1). After that numbers began to decline rapidly through 2003 and 2004, probably linked to legislative changes and then remained at around 4,000 up to 2008. In 2009 numbers fell dramatically to 2,689, a 30% drop on 2008. Coincidentally, over the past 15 years Ireland has seen its economy go from low to high and 'tiger' levels of growth and then back to low growth and recession. Evidently, the existence of these coincidental trends suggests the possibility of a link or level of causality between economic well being and asylum application numbers.

In this paper we conduct a country by country analysis to establish a link between economic well being and asylum application numbers for Ireland and other EEA countries. Specifically we examine whether economic well being as measured by per capita Gross Domestic Product (GDP) volumes in Purchasing Power Standards (PPS) is a factor determining asylum application numbers in Ireland and other EEA countries. For many EEA countries this link is demonstrated showing that the expectation of better incomes is a key reality underpinning asylum movements.

On a purely technical point Gross National Product (GNP) is the preferred measure of a country's 'economic attractiveness'. However, historic data series of per capita GNP volumes in PPS are readily not available. Therefore we opt for per capita GDP volumes in PPS as our measure of economic well being. Crucially this includes net income flows such as repatriated profits made by multinationals which for Ireland are substantial.

Clearly, if economic well being is a predictor of asylum flows then those countries with highest relative incomes should attract more asylum seekers. We examine this by comparing the ratio of per capita asylum applications and per capita

GDP in PPS across EEA countries. Our analysis is based on smoothed ratios obtained by adopting a linear time-varying parameter model to describe this relationship. On this basis we highlight those countries that take a greater share of asylum applicants in terms of their GDP.

Our focus then turns to questions of causality. We look for unit roots and cointegration in the time series and test for the presence of Granger-causality using an error correcting model. Then we proceed to investigate the exact nature of the relationship between per capita asylum application numbers and per capita GDP in PPS in the previous year. Here we assume a multiplicative model for the relationship and once again adopt a time-varying parameter growth model in the logs to describe it. The model is put in state-space form and the parameters themselves are estimated with the aid of the Kalman Filter, see Harvey (1993).

The movement described by the time-varying parameters that results from fitting the model is analysed to see if there is a significant relationship between asylum applications and GDP. The evolution of the parameter also allows us to assess the nature of that relationship through time. Our model also incorporates changing variance and so we are able to see whether uncertainty about economic well being in the host country influences asylum applications. The underlying idea here is that greater uncertainty about future economic conditions will negatively impact on asylum movements and so cause applications to fall.

This paper is structured as follows; in the next section we review our data sources. In Section 3 we look at asylum application trends and loosely cluster EEA countries accordingly. We follow this by modelling the ratio of per capita asylum applications and per capita GDP in PPS to see if there is a difference between countries asylum applications in terms of their GDP. In Section 5 we adopt a constant elasticity model to describe the relationship between asylum applications and GDP and use it to examine issues relating to Granger-causality in Section 6. Finally we use the multiplicative growth model to examine the nature of the relationship between GDP and asylum applications in each EEA country. This model and the associated methodology are described in Sections 7 while conclusions are set out in Section 8.

## 2. Data Sources

Three main sources of data are used for conducting the analysis in this paper. These are:

- Yearly asylum application numbers covering 1985 or later to 2009: http://epp.eurostat.ec.europa.eu/portal/page/portal/population/data/database

- Population number in each year: http://epp.eurostat.ec.europa.eu/portal/page/portal/population/data/database

- Yearly GDP in PPS (constant volume): Gapminder.org: The data in this dataset is based on GDP per capita, in fixed 2005 prices, and is adjusted for Purchasing Power Parities (PPPs), as calculated in the 2005 round of the International Comparison Program (ICP) taken from the Penn World Tables Version 6.2.
  This data source is augmented via regression with GDP per capita for more recent years from Eurostat

http://epp.eurostat.ec.europa.eu/portal/page/portal/national_accounts/data/database

For the purposes of this analysis GDP and where appropriate asylum applications are standardised by dividing each by the overall population giving per capita GDP and per capita asylum applications per 10,000 population respectively. All data are further standardised by dividing by the EU-27 average multiplied by 100.

The data series are short. In Ireland's case they cover 1992-2009 with reliable asylum figures only available from 1995 onward. For this reason our comparisons are made based on data and results over the period from 1997 or 1998 onward - the reason for this is that we drop the first few estimated values obtained by the models to allow for run-in.

## 3. Asylum application trends in EEA countries

Map 1 provides a relative comparison of per capita asylum applications made in Ireland with per capita asylum applications made in other EEA countries. Countries are coloured based on the proportion of years in the 13 year period 1997-2009 where the per capita asylum applications was significantly different to Ireland's (i.e. the proportion of years where per capita asylum applications in that country were statistically not equal to Ireland).

A straightforward binomial test of each country's proportion loosely clusters Austria, Norway, Sweden and Switzerland into a set that have significantly higher asylum applications per head than Ireland. Belgium, Luxembourg and the Netherlands have similar asylum patterns to Ireland while the remaining countries including UK, France, Germany, Italy and Spain comprise a group that have lower asylum applications per head than Ireland. Cyprus, Greece and Malta however have a different pattern to these three groups with high levels of application only since 2004.

## Map 1: per capita Asylum applications in EEA countries relative to Ireland

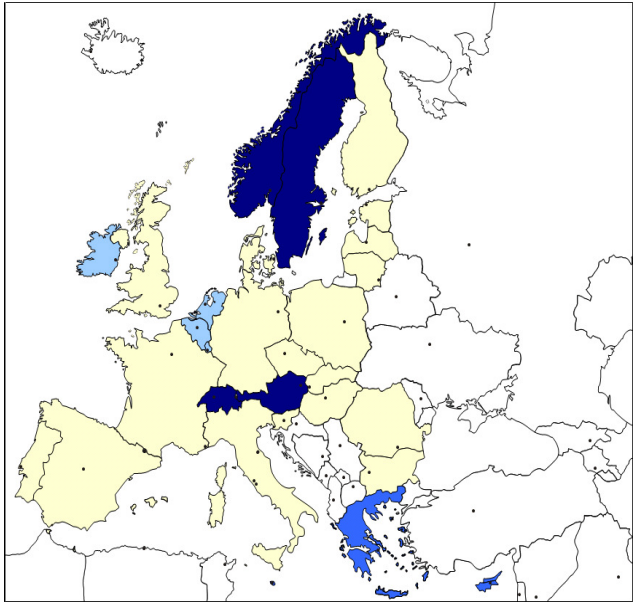

Computing the average of the ratios over the period 1997-2009 (see Appendix Table A1) we find that for Austria, Norway, Sweden and Switzerland the odds of an asylum application being made there is nearly double that of an application being made in Ireland on a per capita basis. For the group with lower per capita asylum applications compared to Ireland (i.e. including UK, France and Germany) the odds of asylum application being made in Ireland is generally over twice that of any one of these countries.

The graphs displayed in Figure 2 are a sample of 'typical' asylum applications per 10,000 of population (all plots are standardised to have a mean and standard deviation of 0 and 1 respectively). Ireland's asylum data only becomes important after the mid-nineties. It is clear that from that point onward a similar pattern is observed to Holland which is in the same group (i.e. having a similar per capita application level), except that applications in Ireland tend to lag Dutch applications by about 2 years. The rise to a peak in the period 1998 to 2002 followed by a rapid fall is also clearly evident from the other plots.

All countries plotted show a sharp rise to a peak in asylum applications in the early nineties; these trends are typical of the application profiles in most EEA states. Germany also displays the early nineties peak but in this instance there is a sustained declined from about 1994 onward - reasons for this provided privately to the author include the introduction of tighter asylum procedures in the nineties. The subgroup showing higher asylum applications than Ireland which includes Switzerland typically also show a new rise from about 2006 or 2007 onward that is not evident for other countries. Interestingly, the timing of two largest asylum application peaks for Switzerland tend to lead those of most other countries by a year or two - this may be a leading indicator of increases in asylum applications in other EEA states.

**Figure 2**

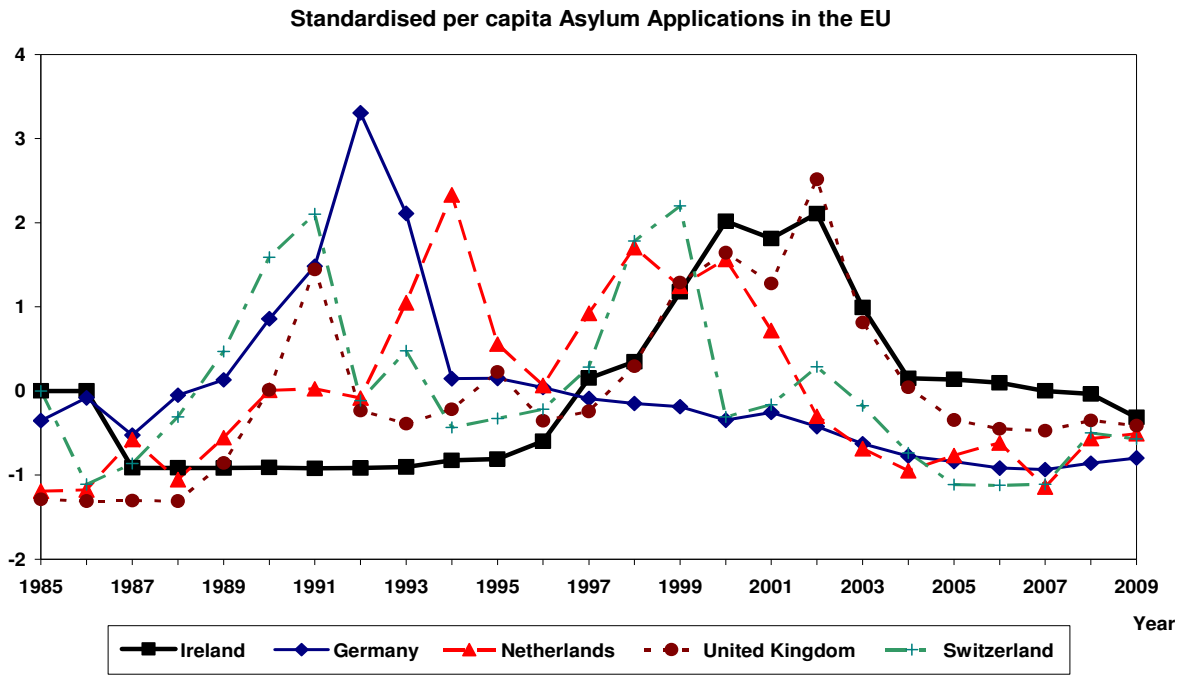

## 4. Asylum application to GDP in PPS ratios in EEA countries

Clearly, if economic well being is related to asylum flows then those countries with highest relative incomes should attract more asylum seekers per capita. This in turn should mean that there is a regional distribution of asylum applications with countries in the 'core' urban so-called 'developed box' that runs from London to Berlin and Stockholm to Milan, see Patrakos et. al. (2005) attracting more applications relative to their GDP than 'periphery' countries. In this section we examine whether there is a regional basis for asylum applications based on each EEA country's GDP.

Our approach is to compare the ratio of per capita asylum application and per capita GDP in PPS across countries and directly identify those countries that take a greater share of asylum applicants in terms of their GDP. Of course for these quantities long-run non-zero linear lag cross-correlations exist but analysis shows there is little evidence for short term stochastic cross-correlations. According to Harvey (1990) characterising the exact nature of this type of non-stationary relationship when data series are short, as is the case here with 25 or fewer observations, is typically best accomplished using an appropriate state-space model. Adopting this approach, if we denote each time point (i.e. year) by the time variable $t$ and label per capita asylum application numbers by $Y_t$ and per capita GDP in PPS by $X_t$, then the crude ratio of asylum to GDP, denoted by $r_t$ is simply

$$r_t = \frac{Y_t}{X_t} \qquad (1)$$

(note: the population element cancels out in the above calculation).

Our preference is to work with smoothed ratios, so using this as our basis we can define a simple time-varying parameter regression model (through the origin) for the smoothed ratio $\rho_t$, expressed in state-space form as

Measurement Eqn:

$$y_t = [x_{t-1}][\rho_t] + e_t \qquad (2a)$$

$$\varepsilon_t \sim N\left(0, \sigma_e^2\right)$$

Transition Eqn:

$$(\rho_t) = [1](\rho_{t-1}) + (n_t) \qquad (2b)$$

$$n_t \sim N\left(0, \sigma_\rho^2\right)$$

Here the smoothed ratio $\rho_t$ is allowed to vary according to a random walk with the variance of this process $\sigma_n^2$ assumed different and independent of the variance of the asylum process $\sigma_e^2$.

The time-varying parameters in this model are estimated in R, Ihaka and Gentleman (1996) using a Kalman Filter, Harvey (1993) and the variance parameters are the maximum likelihood estimates of $\sigma_e^2$ and $\sigma_\rho^2$ available from the prediction error decomposition of the likelihood obtained via the Kalman Filter, see Harvey (1993). These were estimated using the Optim function in R using appropriate starting values. In Figure 3 we provide an example plot of the actual crude ratio $r_t$ and the smoothed ratio $\rho_t$ for the Netherlands. The smoothing is clear from the plot with the peaks and troughs being undershot as the excess noise is damped down at these time points.

**Figure 3: Plot of crude and smoothed asylum to GDP ratios for the Netherlands**

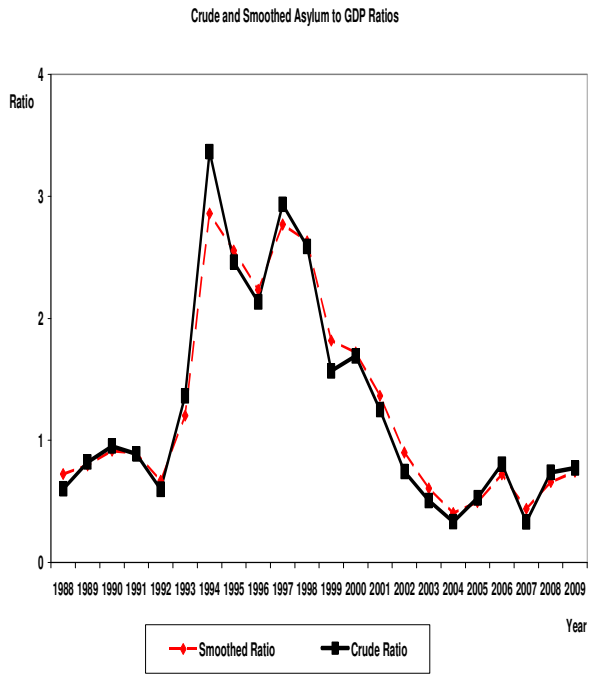

fewer asylum applications in relation to their income. Interestingly, this group also comprises most of the former Eastern European block. For these countries the applications per head is just under half that of Ireland's. Luxembourg, the Netherlands, Norway and Slovakia have similar per capita application levels to Ireland when compared to their wealth. The Eastern Mediterranean group also have much higher applications per head but here high ratios persist only since 2004.

**Map 2: Smoothed per capita Asylum applications to GDP ratios in EEA countries relative to Ireland**

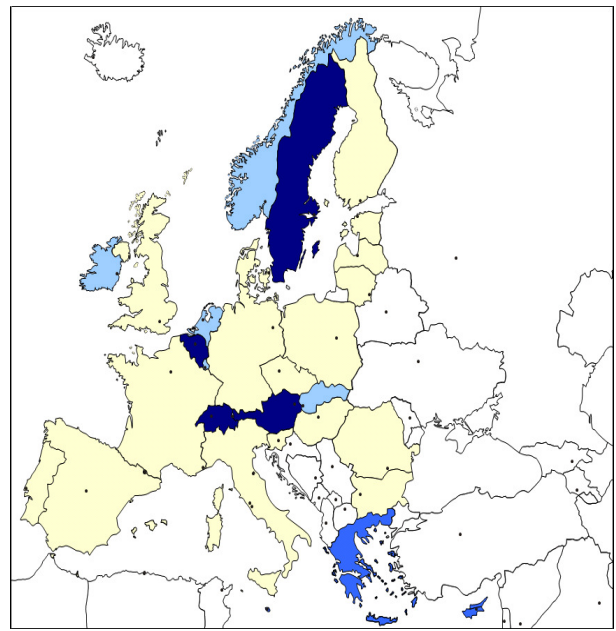

Map 2 shows a relative comparison of the smoothed per capita asylum applications to per capita GDP in PPS ratios in Ireland and in other EEA countries. Once again the colouring reflects the proportion of years (1997-2009) where this ratio was statistically not equal to one (i.e. the per capita asylum applications to GDP ratio in a particular country in that year was different to Ireland).

Statistical analysis of the ratios based on a straightforward binomial test forms 4 simple groups. The first of these includes Austria, Belgium, Sweden and Switzerland. These comprise a group that have significantly higher asylum applications in relation to their national income/wealth per head than Ireland. Based on the average ratio over the time period (see Appendix Table A2) these countries typically have in excess of 1.8 times as many applications per head in relation to their wealth. The bulk of European countries including UK, Denmark, France, Germany, Italy and Spain tend to have

This analysis of per capita asylum applications in terms of national income/wealth shows that Austria, Belgium, Sweden and Switzerland are the only countries in the 'developed box' that take a larger relative proportion than Ireland. Luxembourg and the Netherlands are the only countries in the 'developed box' that take a similar number of applications in relation to their wealth as Ireland. While the UK, Denmark, France and Germany take significantly fewer applications compared to Ireland. Clearly there is no regional pattern to asylum applications in terms of national wealth and the notion that better off countries, i.e. those in the 'developed box', are more attractive to asylum seekers is partially true but cannot be sustained across all developed regions. Additional factors must therefore contribute to explain the regional distribution of asylum applications. A particularly striking example is asylum applications in Germany where the figures in Table A2 (see Appendix) show a steady decline since 1998 - as mentioned above, reasons for this provided privately to the author include the introduction of tighter asylum procedures.

## 5. A constant elasticity model for Asylum and GDP

In this section we hypothesis that there is a constant elasticity or multiplicative growth relationship between per capita asylum application numbers and per capita GDP in PPS described by the following equation:

$$Y_t = \exp^{\mu} \times (X_t)^{\omega} \qquad (3)$$

This equation states that the average per capita asylum application level $(Y)$ is a polynomial function of per capita GDP in PPS $(X)$ or equivalently the relationship is log-log. Here also $\mu$ is a constant and $\omega$ is a constant parameter representing the elasticity between per capita asylum application numbers and per capita GDP in PPS. Thus we assume that the ratio of the relative growth rate between per capita asylum application numbers and per capita GDP in PPS is constant. Clearly, this model is restrictive and neglects other variables that may influence asylum application. Allowing for this, in this paper our focus is on studying the relationship in equation (3).

**Figure 4: Plot of the log of asylum versus GDP for Ireland**

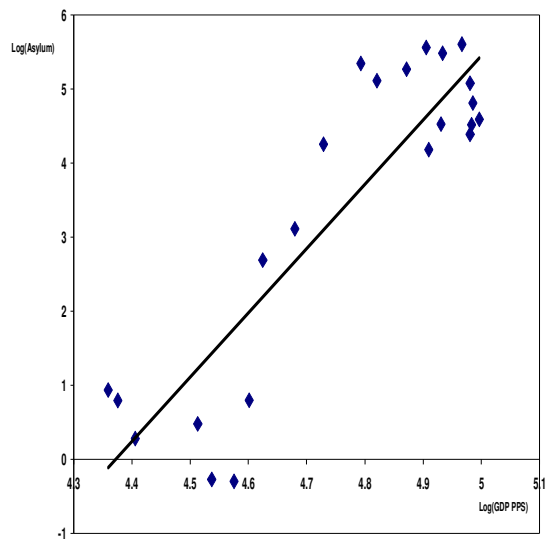

Conceptually we might say that (3) express that GDP expansion leads to higher incomes relative to an asylum seekers home country and this is reflected in relative increases in asylum seeker preferences and causes growth in asylum seeker numbers in the host country. Notwithstanding this, the relationship in (3) is justified on empirical grounds. Taking logs of (3) we have

$$y_t = \mu + \omega x_t \qquad (4)$$

where $y_t = \log(Y_t)$ and $x_t = \log(X_t)$. In Figure 4 the plot of these two variables for Ireland shows a strong linear tendency providing evidence for constant elasticity. That said, it is also clear that there is a tendency for values to fall below or above the trend line in clusters suggesting the errors may not be independent or identically distributed.

## 6. Unit roots, Cointegration and Granger-causality

In this section we examine the nature of the linear dynamic relationship expressed in (4) between logged per capita asylum application numbers and logged per capita GDP in PPS for each individual EEA country. We follow Narayan and Smyth (2004) and use a three stage procedure. First we examine whether these two variables are integrated using the Augmented Dickey-Fuller test available in the R language, Ihaka and Gentleman (1996). The results of these tests at the 5% significance level are reported in Table A3a. For the most part results are as expected as they show that both per capita asylum applications and per capita GDP in PPS are integrated order 1. However, for Latvia, Lithuania and Slovenia the asylum applications variable is stationary while more strikingly for Cyprus, Estonia, Germany, Greece and Poland per capita GDP in PPS appears stationary. Of these the conclusion of the test for Germany would conflict most with expectations. We have also conducted unit root test using the Elliot Rothenburg and Stock (1996) DF-GLS test available in the R language, Ihaka and Gentleman (1996). This test has greater power in small samples. Table 3b gives the results which show that asylum applications in all but 3 countries are stationary. The results for GDP are largely similar to those obtained using the ADF test.

The second step in our examination involves testing for long-run equilibrium relationships between per capita asylum applications and per capita GDP in PPS in logs. We employ the Autoregressive Distributed Lag (ARDL) bound testing procedure of Pesaran, Shin & Smith (2001). This procedure has a few attractive features including that it allows cointegration relationships to be estimated by OLS, the approach is applicable whether the regressors in the model are stationary or integrated order 1, and the test is relatively more efficient in small sample sizes. The ARDL approach is employed by computing the F-statistic based on comparing

the following conditional error correction representations:

$$\Delta y_t = c_0 + \alpha x_{t-1} + \beta y_{t-1} + \sum_{i=1}^{p-1} \psi_i \Delta y_{t-i} + \sum_{i=1}^{p-1} \varphi_i \Delta x_{t-i} + \phi \Delta x_t + u_t \quad (5a)$$

$$\Delta y_t = c_0 + \sum_{i=1}^{p-1} \psi_i \Delta y_{t-i} + \sum_{i=1}^{p-1} \varphi_i \Delta x_{t-i} + \phi \Delta x_t + u_t \quad (5b)$$

where $\Delta$ is the usual 1st-difference operator, $c_0, \psi_i, \varphi_i, \phi$ are regression constants, $p = 2$ is the chosen lag deemed appropriate for annual data and $u_t \sim N\left(0, \sigma_u^2\right)$. The computed F-statistic is compared to the bounds given in Pesaran, Shin & Smith (2001) where three outcomes are specified. First, where the F-statistics is below the lower bound we accept the null hypothesis of no cointegration. Second, where the F-statistic is above the upper bound we reject the null hypothesis in favour of cointegration. Third, the F-statistics lies within the bound and the procedure leaves us undecided. In our case (5) includes a constant and no time trend so the bounds that apply are $3.79 - 4.85$.

The results of the bounds testing procedure are also given in Table A3. For countries where both per capita asylum and per capita GDP in PPS are integrated order 1, these results show that the hypothesis of no cointegration is rejected for the Czech Rep., Finland, Luxembourg, Portugal, UK and Switzerland. Thus for these countries a long run relationship exists between per capita asylum and per capita GDP in PPS when asylum is the dependent variable. The test was inconclusive for France, Italy and Norway. While for Ireland, Austria, Belgium, Bulgaria,

Denmark, Netherlands, Romania, Spain and Sweden there was no evidence for cointegration.

The final step in our test procedure is to examine for Granger-causality. To do this we employ the following pair of ECMs:

$$\left(1-L_{(x)}\right)x_t = \; c_x + \left(1-L_{(y)}\right)\sum_{i=1}^{p-1}\psi_{i(y)}y_{t-i} +$$
$$\left(1-L_{(x)}\right)\sum_{i=1}^{p-1}\varphi_{i(x)}x_{t-i} + u_t \tag{6a}$$

$$\left(1-L_{(y)}\right)y_t = \; c_y + \left(1-L_{(y)}\right)\sum_{i=1}^{p-1}\psi_{i(y)}y_{t-i} +$$
$$\left(1-L_{(x)}\right)\sum_{i=1}^{p-1}\varphi_{i(x)}x_{t-i} + \alpha ECT_{t-1} + u_t \tag{6b}$$

In this equation $L_{(x)}$ and $L_{(y)}$ are the lag operators appropriate for each country per capita GDP in PPS and per capita asylum respectively while $ECT_{t-1}$ is an error correcting term included to account for long-run equilibrium relationships. We test for causality in the ECMs (6) by allowing the lag $p$ to be in the range 1 to 4 and identifying unidirectional causality from Asylum $\rightarrow$ GDP, or GDP $\rightarrow$ Asylum and bidirectional causality between Asylum $\leftrightarrow$ GDP based an the F-probability. For example in testing for causality from Asylum $\rightarrow$ GDP the F-probability compares the full model to a restricted model that does not include lagged $y_{t-i}$ terms.

The results of the causality tests are given in Table A3a. First, when the $ECT_{t-1}$ term was included it was significant at the 10% level for all countries except Slovakia where cointegration was found in the Bounds Test.

Second, bidirectional causality is evident for 8 countries including Ireland, Estonia, Finland, Italy, Malta, Poland, Portugal, Spain and Switzerland. Unidirectional causality from Asylum $\rightarrow$ GDP is evident for Austria, Cyprus, Hungary, Luxembourg, Slovakia and Slovenia. Unidirectional causality from GDP $\rightarrow$ Asylum is evident from Belgium, Czech Rep., France, Sweden and the UK while the remaining countries show no evidence of causality.

Third, in general for countries that showed causality from Asylum $\rightarrow$ GDP either as bi or unidirectional, the lag where causality was found was either 1 or 2 years previously. For the most part the same lag applied for countries that showed causality from GDP $\rightarrow$ Asylum. Causality at short lags is direct evidence for the short-term or stochastic relationship between per capita asylum and per capita GDP in PPS. Interestingly though the UK and Sweden buck this trend with causality showing at lag 4 hinting that it might be necessary to include a time trend in (6) or that parameter estimates over the period may be unstable.

It is also interesting to note that when the causality results in Table 3a are compared with those of Table 3b based on the Elliot Rothenburg Stock unit root test the results are remarkably similar. We are however aware that the absence of cointegration for some countries, including Ireland, based on Elliot Rothenburg Stock test results calls into questions the evidence for causality. Furthermore, we have used standard F-probabilities and in cases where both variables are cointegrated the test statistics may not have a standard F-distribution.

## 7. Asylum and GDP elasticity and predictability in EEA countries

In this section we model the dynamic relationship between per capita asylum application numbers and per capita GDP in PPS for each individual EEA country. Rather than use the ECM framework adopted in the previous section, we model the relationships directly under the assumption that the elasticity is not a fixed constant but can vary randomly. Modelling the relationship directly has the advantage we can work with non-stationary data. It also allows us to directly compute the elasticity between GDP and asylum numbers to see if it is similar or different across the EEA. Indeed, the Granger-causality results of the previous section suggest we should focus our attention on the elasticity at short lags. So, we specifically look at the elasticity between last year's GDP this year's asylum numbers.

To investigate these issues we assume that the following non-stationary time-varying parameter model describes the multiplicative growth relationship between per capita asylum application numbers and per capita GDP in PPS:

$$Y_t = \exp^{\mu} \times (X_{t-1})^{\omega_t} \qquad (7)$$

This equation states that the average per capita asylum application level $(Y)$ is a polynomial function of per capita GDP in PPS $(X)$ in the previous year. Here also $\exp^{\mu}$ is a constant and $\omega_t$ is a time-varying parameter representing the time varying elasticity between per capita asylum application numbers and per capita GDP in PPS.

Taking logs we can readily express this model in state space form as:

Measurement Eqn:

$$y_t = \begin{bmatrix} 1 & x_{t-1} \end{bmatrix} \begin{bmatrix} \mu_t \\ \omega_t \end{bmatrix} + \varepsilon_t \qquad \varepsilon_t \sim N(0, \sigma_\varepsilon^2) \qquad (8a)$$

Transition Eqn:

$$\begin{pmatrix} \mu_t \\ \omega_t \end{pmatrix} = \begin{bmatrix} 1 & 0 \\ 0 & 1 \end{bmatrix} \begin{pmatrix} \mu_{t-1} \\ \omega_{t-1} \end{pmatrix} + \begin{pmatrix} 0 \\ \eta_{2t} \end{pmatrix} \qquad (8b)$$

$$\eta_{2t} \sim N(0, \sigma_\omega^2)$$

where $y_t = \log_e(Y_t)$ and $x_{t-1} = \log_e(X_{t-1})$. Importantly, this model allows the elasticity to vary according to a random walk. The variance or uncertainty of this process $\sigma_\eta^2$ is assumed different and independent of the variance of the asylum process $\sigma_e^2$ given in equation (8a). Once again we implement a Kalman Filter, see Harvey (1993) in the R language, Ihaka and Gentleman (1996) to obtain the smoothed time-varying parameters while the variance parameters are the maximum likelihood estimates of $\sigma_e^2$ and $\sigma_\omega^2$ available from the Kalman Filter. These were estimated using the Optim function in R. All our model results given below are obtained from fitting this model to each country's data.

**Results**

Table 1 shows the results obtained when the time varying parameter model was fitted to each country's data. In the table $n$ is the number of observations while $\hat{\sigma}_e^2$, $\hat{\sigma}_\mu^2$ and $\hat{\sigma}_\omega^2$ are the estimated variances of the residual, the constant and time varying elasticity respectively. The filtered constant $\hat{\mu}$ and the average value of filtered elasticity $\overline{\omega}_t$ are also given. Where this parameter is not significantly different from 0 the country is shown shaded. Note that $\mu$ is initially assumed stochastic and the model is re-estimated as in equation (8) when this parameter is not significantly different from 0 at the 5% level.

The next two columns in Table 1 are the sum of the Absolute Residual Errors (ARE) and the Sum of Squares of the Errors (SSE), both expressed as a percentage of the corresponding sum of the per capita asylum applications. The column labelled Box-Pierce, Harvey (1990) is the outcome of the Box-Pierce test - where the null hypothesis is that there is no autocorrelation in the residuals. The last column labelled ARCH shows whether the conditional variance is statistically significant. Here 'Accept' means that the variance and therefore uncertainty about future expectations changes over time. We note that the conditional variances of the model are available for free from the Kalman Filter, see Kim & Nelson (1999, p45).

In Table 1 we can see the model has tracked the data for each country closely. This is evidenced by the fact that the estimated residual variance $\hat{\sigma}_e^2$ is small in most cases and the ARE and SSE are both small. In tandem with the error being small the Box-Pierce test shows that no autocorrelation was observed in the residuals. Therefore, the model provides a good representation of the data.

It is also clear from Table 1 that $\hat{\sigma}_\mu^2 = 0$ in all cases except for Luxembourg, Spain and Slovenia. Thus the average level of asylum applications $\hat{\mu}$ is constant for all but these three countries. Moreover, the uncertainty associated with the estimated average levels is large and therefore these quantities, whether constant or time varying, are not statistically different from 0.

Of primary interest is the filtered elasticity $\overline{\hat{\omega}}_t$. For all countries this quantity is positive while the variance of the slope is $\hat{\sigma}_\omega^2 > 0$ in all cases except for Luxembourg, Spain and Slovenia.

Therefore the assumed time varying multiplicative growth model captures the dynamic relationship between per capita GDP and per capita asylum applications for most EEA countries. For countries in the bottom group in the table the standard errors tend to include 0 and therefore inferences are less reliable. This group mainly comprise the eastern bloc within the EU, a reflection of the fact that the data series are extremely short for these countries.

Importantly, for the 16 countries where reliable inferences can be made we can conclude that asylum applications depend on the previous years GDP. In particular, in Ireland's case since $\overline{\hat{\omega}}_t = 1.09$ we can say that a 1% increase in per capita GDP in PPS in any given year will produce 1.09% increase in per capita asylum applications in the following year. For other countries the response varies from the lowest group, where 1% increase in per capita GDP in PPS in any given year will produce 0.5% increase in per capita asylum applications in the following year, while the response in the UK is about 0.96%, slightly lower than Ireland's.

Table 1: Filtered Parameter Values from fitting the time-varying parameter model

| Country | $n$ | $\hat{\sigma}_e^2$ | $\hat{\sigma}_\mu^2$ | $\hat{\sigma}_\omega^2$ | $\hat{\mu}$ | $\overline{\omega}_t$ | ARE (%) | SSE (%) | Box-Pierce (Accept) | ARCH (Accept) |
|---|---|---|---|---|---|---|---|---|---|---|
| | | MLEs of the variances | | | Time varying parameters | | Model Quality/Accuracy Tests | | | |
| Cyprus | 12 | 0.23 | 0.00 | 0.13 | 0.33 | 1.28 | 5.5 | 0.4 | Y | Y |
| Sweden | 25 | 0.22 | 0.00 | 0.05 | 0.08 | 1.14 | 11.2 | 2.1 | Y | Y |
| Ireland | 15 | 0.25 | 0.00 | 0.05 | -0.07 | 1.09 | 8.3 | 1.1 | Y | N |
| Norway | 25 | 0.17 | 0.00 | 0.09 | -0.23 | 1.08 | 4.6 | 0.4 | Y | Y |
| Switzerland | 24 | 0.18 | 0.00 | 0.07 | 0.28 | 1.06 | 5.1 | 0.3 | Y | N |
| Belgium | 25 | 0.14 | 0.00 | 0.07 | 0.19 | 1.04 | 4.8 | 0.3 | Y | N |
| Austria | 25 | 0.18 | 0.00 | 0.07 | 0.21 | 1.03 | 4.7 | 0.3 | Y | N |
| Malta | 13 | 0.23 | 0.00 | 0.10 | 0.59 | 1.03 | 6.4 | 0.6 | Y | Y |
| Denmark | 25 | 0.20 | 0.00 | 0.06 | -0.03 | 1.00 | 8.3 | 0.9 | Y | N |
| United Kingdom | 25 | 0.22 | 0.00 | 0.08 | -0.20 | 0.96 | 6.9 | 0.8 | Y | N |
| Netherlands | 25 | 0.24 | 0.00 | 0.08 | 0.12 | 0.95 | 9.2 | 1.7 | Y | N |
| Slovakia | 18 | 0.28 | 0.00 | 0.17 | 0.01 | 0.93 | 6.1 | 0.8 | Y | Y |
| Poland | 16 | 0.35 | 0.00 | 0.06 | -0.67 | 0.89 | 15.6 | 4.4 | Y | N |
| France | 19 | 0.14 | 0.00 | 0.04 | 0.04 | 0.86 | 5.6 | 0.4 | Y | N |
| Germany | 25 | 0.23 | 0.00 | 0.06 | 0.24 | 0.83 | 8.2 | 0.7 | Y | N |
| Finland | 25 | 0.29 | 0.00 | 0.13 | -0.26 | 0.79 | 6.9 | 1.1 | Y | Y |
| Greece | 25 | 0.29 | 0.00 | 0.12 | 0.54 | 0.74 | 6.9 | 1.1 | Y | N |
| Czech Republic | 13 | 0.22 | 0.00 | 0.10 | 0.56 | 0.72 | 8.2 | 1.1 | Y | N |
| Hungary | 12 | 0.28 | 0.00 | 0.12 | 0.57 | 0.69 | 7.5 | 0.5 | Y | N |
| Luxembourg | 25 | 0.24 | 0.51 | 0.00 | 0.97 | 0.67 | 7.4 | 0.8 | Y | N |
| Spain | 25 | 0.13 | 0.26 | 0.00 | 0.10 | 0.60 | 4.6 | 0.4 | Y | N |
| Bulgaria | 13 | 0.22 | 0.00 | 0.09 | 0.69 | 0.55 | 7.0 | 0.6 | Y | N |
| Italy | 25 | 0.66 | 0.00 | 0.94 | 0.37 | 0.55 | 29.1 | 14.2 | Y | Y |
| Lithuania | 13 | 0.27 | 0.00 | 0.14 | 0.68 | 0.32 | 8.5 | 1.6 | Y | N |
| Latvia | 12 | 0.56 | 0.00 | 0.14 | -0.24 | 0.30 | 19.4 | 6.2 | Y | N |
| Portugal | 25 | 0.26 | 0.00 | 0.13 | -0.56 | 0.30 | 9.7 | 2.8 | Y | N |
| Slovenia | 16 | 0.24 | 0.97 | 0.00 | 2.21 | 0.28 | 8.8 | 1.6 | Y | N |
| Romania | 19 | 0.11 | 0.00 | 0.11 | 0.70 | 0.23 | 2.5 | 0.1 | Y | N |
| Estonia | 12 | 0.28 | 0.00 | 0.09 | -0.64 | 0.11 | 16.1 | 5.6 | Y | N |

| Table 2: Actual smoothed per capita Asylum applications to GDP elasticises 2006-2009 | | | | | | | |
|---|---|---|---|---|---|---|---|
| **Year** | **Ireland** | **Austria** | **Belgium** | **Cyprus** | **Denmark** | **Finland** | **France** | **Germany** |
| **2006** | 0.98 | 1.01 | 0.95 | 1.38 | 0.94 | 0.90 | 0.88 | 0.63 |
| **2007** | 0.96 | 0.95 | 1.41 | 1.29 | 0.93 | 0.78 | 0.90 | 0.57 |
| **2008** | 0.95 | 0.96 | 1.02 | 1.29 | 0.94 | 0.94 | 0.88 | 0.62 |
| **2009** | 0.95 | 1.00 | 1.09 | 1.22 | 0.98 | 1.05 | 0.89 | 0.68 |
| **Year** | **Malta** | **Netherlands** | **Poland** | **Slovakia** | **Sweden** | **UK** | **Norway** | **Switzerland** |
| **2006** | 1.16 | 0.91 | 0.81 | 0.98 | 1.16 | 0.89 | 1.00 | 1.00 |
| **2007** | 1.15 | 0.80 | 0.87 | 0.90 | 1.21 | 0.87 | 1.01 | 0.98 |
| **2008** | 1.27 | 0.89 | 0.95 | 0.60 | 1.19 | 0.88 | 1.13 | 1.07 |
| **2009** | 1.28 | 0.92 | 1.05 | 0.54 | 1.19 | 0.89 | 1.18 | 1.08 |

**Map 3: Smoothed per capita Asylum applications to GDP elasticises in EEA countries relative to Ireland**

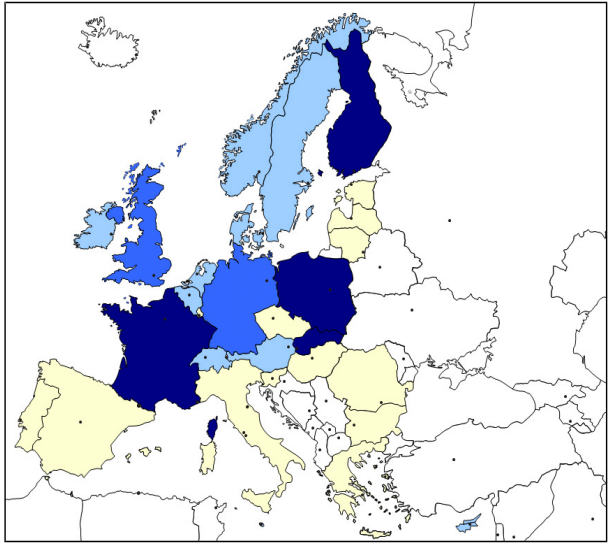

Using Ireland as a baseline, Map 3 (also Appendix Table A3) shows the geographic distribution of elasticity between asylum numbers and GDP. Once again the colouring reflects the proportion of years (1998-2009) where the elasticity relative to Ireland was statistically not equal to one (i.e. the asylum applications to GDP elasticity in a particular country in that year was different to Ireland). Statistical analysis of the elasticity ratios based on a straightforward binomial test forms 3 simple groups. An appealing feature of this simple technique is that the first and second groups include 15 of the 16 countries where reliable inferences can be made about the elasticity. Meanwhile within this 15 only the UK and Germany have elasticises that are significantly lower than Ireland based standard error of their elasticity.

All remaining countries except Greece have a significantly lower response that cannot be distinguished from 0 given the data. In general therefore there is no significant difference and indeed there is remarkable homogeneity across EEA countries in relation to the response of asylum applications to changes in economic well being. This response is always positive in that an increase in GDP produces an increase in asylum applications. Interpreting this in terms of asylum seeker expectations, we may say this indicates that when there is a perception that a particular country's economy is growing, there may be a latent expectation among asylum applicants of a favourable outcome to their asylum claim.

More intriguingly from a long-run perspective we might have expected geographic variation in GDP, see Patrakos et. al. (2005) to be also reflected in asylum application elasticises. Clearly this is not the case. Analysis of elasticises tends to support the view that asylum seekers take it that incomes have largely evened out in the more developed EEA economies – this is the so called neo-classical view of economic convergence. Thus, even though economic growth effects accumulate in a regionally selective manner, see Bradley et. al. (2006) in the EEA and growing economies are more attractive to asylum seekers, the asylum seeker has the same view as the economic migrant. That is, when the relative income in the host country exceeds that in the home country migration occurs.

Clearly in recent years the general economic situation has been is flux. This raises a couple of interesting short-run questions. First, countries experiencing a downturn should see falls in applications while those where the economic conditions remain favourable might see application increase. That is to say a preference for one asylum destination over another may be influenced by economic conditions. In Table 2 we give the actual smoothed per capita asylum applications to GDP elasticises for the years 2006-2009 for the group of 16 countries. It is clear there is an economic preference for asylum seekers for the Nordic countries. In Ireland the economic conditions have got worse over the period while Switzerland has seen its per capita GDP in PPS improve by 5% over the same period. The elasticises reflect this improvement with an increase of 8% from 1.00 to 1.08 occurring in this time period. Given the steady level of per capita asylum applications in most countries this acceleration shows that the economic preferences of new asylum applicants would appear to have recently shifted away from Ireland toward Switzerland. However, it remains to be seen as to whether this effect is permanent. Recently further evidence for this shift has emerged from Nigerian asylum applications. For the past 10 years Nigerians have been the largest nationality group making asylum applications in Ireland. Interestingly in 2009 this number halved while in Switzerland they have noticed a large increase in applications from Nigerians.

The second question that arises on foot of the flux in recent economic conditions is whether uncertainty about future economic expectations has a bearing on asylum applications. The final column of Table 1, labelled ARCH, gives the outcome of the test for conditional variance which is a measure of future uncertainty. For the group of 16 countries the clear picture that arises from this test is that in most central EU member states the uncertainty remains constant over time.

However, it is also clear that there is varying uncertainty in asylum applications in Nordic countries as well as Malta and Cyprus over time in response to uncertainty in economic conditions. Figure 3 plots the uncertainty for Sweden. The plot shows that the variability tends to fall but with a small spike in the early nineties, this is probably due to the effect of the small sample used to compute the early forecasts. It is clear from this that the banking crisis in the early 1990s in Nordic countries had a small but largely insignificant effect on asylum variability in applications. The recent Irish banking crisis seems to mirror this in that Ireland has not experienced greater variability in the number of applications as a consequence of uncertainty about the country's future economic well being.

Figure 3

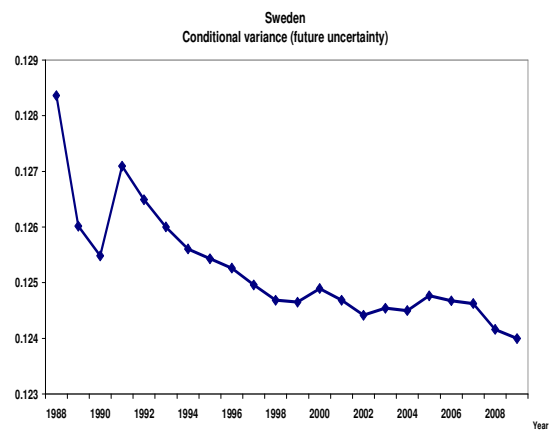

## 8. Conclusions and Observations

In this paper we have looked at asylum application trends across the EEA and related them to GDP. Analysis of per capita asylum applications showed that Ireland, Belgium and the Netherlands appear to be taking a similar number of asylum applications in relation to their population relative to other EEA states. In Austria, Norway, Sweden and Switzerland the odds of an asylum application being made there is nearly double that of an application being made Ireland while in most other countries including UK, France and Germany the odds of asylum application being made there is just ½ that of Ireland's.

Examining the relationship between asylum applications and GDP showed a largely similar pattern as the analysis based on per capita applications alone. Austria, Belgium, Sweden and Switzerland are the only countries in the 'developed box' taking more applications per head in relation to their wealth as Ireland. Meanwhile the UK, Denmark, Finland, France, Germany and Spain have fewer asylum applications in relation to their income. As a consequence the notion that better off countries are more attractive to asylum seekers was shown to be partially true but could not be sustained across all developed regions.

The Granger causality analysis showed that 14 of the 29 countries consider displayed evidence of causality between asylum and GDP. Of these 14 countries 8 are located in Western Europe. Invariably the causality found was short term, at 1 to 2 year lags. Long term relationships existed between asylum and GDP for 8 countries with 6 of these located in Western Europe.

The dynamic analysis of elasticises demonstrated that asylum applications are a function of economic well being in the previous year for a group of 16 countries mainly located in Western Europe. In Ireland's case a 1% increase in per capita GDP in PPS in any given year produces 1.09% increase in per capita asylum applications in the following year.

From a more general perspective the long run analysis showed that asylum seekers tend to see the EEA as economically homogeneous. Nevertheless recent relative growth rates indicate that economic preferences of new asylum seekers appear to have shifted away from Ireland and the UK and toward Switzerland.

In this study we have shown that there is an association and a measurable response in asylum applications to movements in GDP. The relationships were repeated across several EEA countries. Moreover, the long run attraction of the EEA as a whole to asylum seekers is a phenomenon that is analogous to more general migration. Thus the overall impression that arises from this study is that asylum applications are driven in part by GDP but that this contributory causal relationship varies from country to county in the EEA. More recent GDP patterns tend to act as random shocks altering asylum applications levels in the short run.

Clearly, there is more to asylum applications than just GDP. Family ties or looser social networks, as well as legislative and administrative procedures surrounding asylum, are important factors. These topics warrant future research. But there are also other global and geo-political features. For example, Menjivar (1993), in a study on Salvadorian refugee migration to the US, shows that migrations are seldom the result of a sudden crisis, such as a war, but are related to the broad historical process in which political and economic factors interact. This indicates that people become refugees through a decline in their freedoms while others, as we have shown in this study, are simply impelled to move mainly due to economic decline.

Appendix: Table A1 - per capita Asylum applications in EEA countries relative to Ireland (1)

| Country | 1997 | 1998 | 1999 | 2000 | 2001 | 2002 | 2003 | 2004 | 2005 | 2006 | 2007 | 2008 | 2009 | Proportion of years greater than 1 | Average |
|---|---|---|---|---|---|---|---|---|---|---|---|---|---|---|---|
| Ireland | 1 | 1 | 1 | 1 | 1 | 1 | 1 | 1 | 1 | 1 | 1 | 1 | 1 | 0.50 | 1 |
| Belgium | 1.1 | 1.7 | 1.7 | 1.4 | 0.9 | 0.6 | 0.7 | 1.1 | 1.1 | 0.8 | 1.2 | 1.7 | 3.5 | 0.69 | 1.4 |
| Luxembourg | 1.0 | 3.2 | 3.3 | 0.5 | 0.6 | 0.8 | 1.8 | 3.3 | 1.7 | 1.1 | 1.0 | 1.1 | 1.6 | 0.62 | 1.6 |
| Netherlands | 2.1 | 2.3 | 1.2 | 1.0 | 0.8 | 0.4 | 0.4 | 0.6 | 0.7 | 0.9 | 0.5 | 1.1 | 1.6 | 0.38 | 1.0 |
| Austria | 0.8 | 1.4 | 1.2 | 0.8 | 1.4 | 1.6 | 2.1 | 2.9 | 2.6 | 1.6 | 1.6 | 1.7 | 3.1 | 0.85 | 1.8 |
| Norway | 0.5 | 1.5 | 1.1 | 0.8 | 1.2 | 1.3 | 1.9 | 1.6 | 1.1 | 1.1 | 1.5 | 3.5 | 5.9 | 0.85 | 1.8 |
| Sweden | 1.0 | 1.2 | 0.6 | 0.6 | 1.0 | 1.2 | 1.9 | 2.4 | 1.9 | 2.7 | 4.4 | 3.1 | 4.3 | 0.77 | 2.0 |
| Switzerland | 3.2 | 4.5 | 3.0 | 0.9 | 1.0 | 1.1 | 1.4 | 1.7 | 1.2 | 1.2 | 1.4 | 2.5 | 3.4 | 0.85 | 2.0 |
| Bulgaria | 0.0 | 0.1 | 0.1 | 0.1 | 0.1 | 0.1 | 0.1 | 0.1 | 0.1 | 0.1 | 0.1 | 0.1 | 0.2 | 0.00 | 0.1 |
| Czech Republic | 0.2 | 0.3 | 0.3 | 0.3 | 0.7 | 0.3 | 0.6 | 0.5 | 0.3 | 0.3 | 0.2 | 0.2 | 0.2 | 0.00 | 0.3 |
| Denmark | 0.9 | 0.9 | 0.6 | 0.7 | 0.9 | 0.4 | 0.4 | 0.6 | 0.4 | 0.4 | 0.4 | 0.5 | 1.1 | 0.08 | 0.6 |
| Germany | 1.2 | 1.0 | 0.6 | 0.3 | 0.4 | 0.3 | 0.3 | 0.4 | 0.3 | 0.3 | 0.3 | 0.4 | 0.7 | 0.08 | 0.5 |
| Estonia | 0.0 | 0.0 | 0.0 | 0.0 | 0.0 | 0.0 | 0.0 | 0.0 | 0.0 | 0.0 | 0.0 | 0.0 | 0.0 | 0.00 | 0.0 |
| Finland | 0.2 | 0.2 | 0.3 | 0.2 | 0.1 | 0.2 | 0.3 | 0.6 | 0.7 | 0.4 | 0.3 | 0.8 | 1.8 | 0.08 | 0.5 |
| France | 0.3 | 0.3 | 0.2 | 0.2 | 0.3 | 0.3 | 0.5 | 0.9 | 0.8 | 0.5 | 0.5 | 0.7 | 1.2 | 0.08 | 0.5 |
| Italy | 0.0 | 0.2 | 0.2 | 0.1 | 0.1 | 0.1 | 0.1 | 0.2 | 0.2 | 0.2 | 0.3 | 0.6 | 0.5 | 0.00 | 0.2 |
| Latvia | | 0.0 | 0.0 | 0.0 | 0.0 | 0.0 | 0.0 | 0.0 | 0.0 | 0.0 | 0.0 | 0.0 | 0.0 | 0.00 | 0.0 |
| Lithuania | 0.1 | 0.0 | 0.0 | 0.0 | 0.0 | 0.0 | 0.1 | 0.0 | 0.0 | 0.0 | 0.0 | 0.2 | 0.2 | 0.00 | 0.1 |
| Hungary | | 0.6 | 0.5 | 0.3 | 0.3 | 0.2 | 0.1 | 0.1 | 0.2 | 0.2 | 0.4 | 0.4 | 0.8 | 0.00 | 0.3 |
| Poland | 0.1 | 0.1 | 0.0 | 0.0 | 0.0 | 0.0 | 0.1 | 0.2 | 0.1 | 0.1 | 0.2 | 0.3 | 0.5 | 0.00 | 0.1 |
| Portugal | 0.0 | 0.0 | 0.0 | 0.0 | 0.0 | 0.0 | 0.0 | 0.0 | 0.0 | 0.0 | 0.0 | 0.0 | 0.0 | 0.00 | 0.0 |
| Romania | 0.1 | 0.0 | 0.0 | 0.0 | 0.0 | 0.0 | 0.0 | 0.0 | 0.0 | 0.0 | 0.0 | 0.1 | 0.1 | 0.00 | 0.0 |
| Slovenia | 0.0 | 0.1 | 0.2 | 1.6 | 0.3 | 0.1 | 0.3 | 0.5 | 0.7 | 0.2 | 0.2 | 0.1 | 0.2 | 0.08 | 0.4 |
| Slovakia | 0.1 | 0.1 | 0.1 | 0.1 | 0.6 | 0.6 | 1.0 | 2.0 | 0.6 | 0.5 | 0.5 | 0.2 | 0.3 | 0.15 | 0.5 |
| Spain | 0.1 | 0.1 | 0.1 | 0.1 | 0.1 | 0.1 | 0.1 | 0.1 | 0.1 | 0.1 | 0.2 | 0.1 | 0.1 | 0.00 | 0.1 |
| United Kingdom | 0.5 | 0.6 | 0.6 | 0.5 | 0.4 | 0.6 | 0.5 | 0.6 | 0.5 | 0.5 | 0.5 | 0.6 | 0.8 | 0.00 | 0.6 |
| Cyprus | | 0.3 | 0.6 | 0.3 | 0.6 | 0.5 | 3.3 | 12.5 | 6.8 | 5.9 | 9.5 | 5.0 | 5.5 | 0.58 | 4.5 |
| Greece | 0.1 | 0.2 | 0.1 | 0.1 | 0.2 | 0.2 | 0.4 | 0.4 | 0.8 | 1.1 | 2.5 | 2.0 | 2.3 | 0.31 | 0.8 |
| Malta | 0.2 | 0.3 | 0.3 | 0.1 | 0.1 | 0.3 | 0.6 | 2.3 | 2.8 | 3.1 | 3.7 | 7.2 | 9.5 | 0.46 | 2.4 |

**Appendix: Table A2 - Smoothed per capita Asylum applications to GDP ratios in EEA countries relative to Ireland**

| Country | Year | | | | | | | | | | | | | Proportion of years greater than 1 | Average |
|---|---|---|---|---|---|---|---|---|---|---|---|---|---|---|---|
| | 1997 | 1998 | 1999 | 2000 | 2001 | 2002 | 2003 | 2004 | 2005 | 2006 | 2007 | 2008 | 2009 | | |
| Ireland | 1 | 1 | 1 | 1 | 1 | 1 | 1 | 1 | 1 | 1 | 1 | 1 | 1 | 0.5 | 1 |
| Luxembourg | 0.6 | 1.4 | 1.6 | 0.4 | 0.3 | 0.4 | 0.8 | 1.3 | 1.0 | 0.7 | 0.5 | 0.5 | 0.7 | 0.3 | 0.8 |
| Netherlands | 2.0 | 2.0 | 1.3 | 1.0 | 0.8 | 0.5 | 0.5 | 0.6 | 0.7 | 0.9 | 0.7 | 1.0 | 1.4 | 0.3 | 1.0 |
| Norway | 0.4 | 0.8 | 0.8 | 0.6 | 0.8 | 0.9 | 1.3 | 1.3 | 1.0 | 0.9 | 1.2 | 2.3 | 3.7 | 0.4 | 1.2 |
| Slovakia | 0.3 | 0.2 | 0.3 | 0.3 | 1.3 | 1.5 | 2.2 | 3.7 | 1.7 | 1.3 | 1.1 | 0.4 | 0.4 | 0.5 | 1.1 |
| Austria | 0.9 | 1.1 | 1.1 | 0.8 | 1.3 | 1.7 | 2.0 | 2.5 | 2.8 | 2.0 | 1.8 | 1.8 | 2.7 | 0.9 | 1.7 |
| Belgium | 1.3 | 1.5 | 1.7 | 1.5 | 1.1 | 0.8 | 0.8 | 1.0 | 1.3 | 1.1 | 1.3 | 1.8 | 3.3 | 0.8 | 1.4 |
| Sweden | 1.3 | 1.2 | 0.9 | 0.7 | 0.9 | 1.1 | 1.7 | 2.7 | 2.6 | 2.7 | 3.9 | 3.7 | 4.6 | 0.8 | 2.2 |
| Switzerland | 2.9 | 3.4 | 2.8 | 1.2 | 1.0 | 1.1 | 1.3 | 1.6 | 1.4 | 1.3 | 1.4 | 2.2 | 2.9 | 1.0 | 1.9 |
| Slovenia | 0.0 | 0.1 | 0.1 | 0.5 | 0.5 | 0.4 | 0.6 | 1.0 | 1.1 | 0.9 | 1.0 | 0.9 | 1.1 | 0.2 | 0.6 |
| Germany | 2.1 | 1.5 | 1.0 | 0.6 | 0.6 | 0.5 | 0.5 | 0.6 | 0.5 | 0.4 | 0.4 | 0.5 | 0.7 | 0.2 | 0.8 |
| Denmark | 1.0 | 0.8 | 0.6 | 0.6 | 0.8 | 0.5 | 0.5 | 0.6 | 0.5 | 0.4 | 0.5 | 0.5 | 1.0 | 0.1 | 0.7 |
| Finland | 0.2 | 0.2 | 0.3 | 0.3 | 0.2 | 0.2 | 0.3 | 0.6 | 0.8 | 0.6 | 0.4 | 0.8 | 1.6 | 0.1 | 0.5 |
| France | 0.4 | 0.3 | 0.3 | 0.2 | 0.3 | 0.3 | 0.5 | 0.8 | 0.9 | 0.7 | 0.7 | 0.9 | 1.3 | 0.1 | 0.6 |
| Bulgaria | | | | 0.3 | 0.5 | 0.5 | 0.4 | 0.4 | 0.4 | 0.3 | 0.4 | 0.4 | 0.5 | 0.0 | 0.4 |
| Czech Rep. | | | | 0.5 | 1.0 | 0.7 | 1.0 | 1.0 | 0.8 | 0.6 | 0.4 | 0.3 | 0.3 | 0.0 | 0.7 |
| Estonia | 0.0 | 0.0 | 0.0 | 0.0 | 0.0 | 0.0 | 0.0 | 0.0 | 0.0 | 0.0 | 0.0 | 0.0 | 0.1 | 0.0 | 0.0 |
| Hungary | | | | | 0.8 | 0.5 | 0.3 | 0.3 | 0.3 | 0.5 | 0.8 | 0.8 | 1.4 | 0.0 | 0.6 |
| Italy | 0.0 | 0.1 | 0.2 | 0.1 | 0.1 | 0.1 | 0.2 | 0.2 | 0.2 | 0.2 | 0.3 | 0.6 | 0.7 | 0.0 | 0.2 |
| Latvia | | | | | 0.0 | 0.0 | 0.0 | 0.0 | 0.0 | 0.0 | 0.0 | 0.1 | 0.1 | 0.0 | 0.0 |
| Lithuania | | | | 0.1 | 0.1 | 0.1 | 0.2 | 0.1 | 0.1 | 0.1 | 0.1 | 0.4 | 0.5 | 0.0 | 0.2 |
| Poland | 0.2 | 0.2 | 0.1 | 0.1 | 0.1 | 0.1 | 0.2 | 0.4 | 0.4 | 0.3 | 0.5 | 0.6 | 0.9 | 0.0 | 0.3 |
| Portugal | 0.0 | 0.0 | 0.0 | 0.0 | 0.0 | 0.0 | 0.0 | 0.0 | 0.0 | 0.0 | 0.0 | 0.0 | 0.0 | 0.0 | 0.0 |
| Romania | 0.2 | 0.2 | 0.2 | 0.1 | 0.1 | 0.1 | 0.1 | 0.1 | 0.1 | 0.1 | 0.1 | 0.2 | 0.2 | 0.0 | 0.1 |
| Spain | 0.2 | 0.1 | 0.1 | 0.1 | 0.1 | 0.1 | 0.1 | 0.1 | 0.2 | 0.2 | 0.2 | 0.2 | 0.2 | 0.0 | 0.1 |
| UK | 0.6 | 0.6 | 0.6 | 0.5 | 0.5 | 0.6 | 0.7 | 0.8 | 0.7 | 0.6 | 0.6 | 0.7 | 0.8 | 0.0 | 0.6 |
| Greece | 0.6 | 0.4 | 0.4 | 0.1 | 0.2 | 0.3 | 0.5 | 0.5 | 1.0 | 1.5 | 3.0 | 2.8 | 2.9 | 0.2 | 1.1 |
| Cyprus | | | | | 1.1 | 0.7 | 3.6 | 12.7 | 14.7 | 10.2 | 13.7 | 8.5 | 7.6 | 0.0 | 6.1 |
| Malta | | | | 0.3 | 0.2 | 0.5 | 0.9 | 3.0 | 4.8 | 5.8 | 6.6 | 11.3 | 14.5 | 0.7 | 4.8 |

**Appendix: Table A3a – ADF Unit root, Cointegration and Granger-causality test results**

| Country | N | ADF Unit Root Test Integration order $\alpha = 0.05$ | | Bounds Cointegration Test H₀: No Cointegration F statistic $\alpha = 0.05$ | | Causality Direction | | | | Error correction term (significance level) |
| | | | | | | Asylum -> GDP | | GDP -> Asylum | | |
| | | Asylum | GDP | Accept < 3.79 Reject > 4.85 | Test Result | F-prob. | Lag | F-prob. | Lag | |
|---|---|---|---|---|---|---|---|---|---|---|
| Ireland | 12 | 1 | 1 | 1.23 | Accept | 0.94 | 1 | 0.95 | 3 | N/A |
| Austria | 22 | 1 | 1 | 2.96 | Accept | 0.91 | 1 | 0.18 | 1 | N/A |
| Belgium | 22 | 1 | 1 | 1.94 | Accept | 0.86 | 2 | 0.92 | 1 | N/A |
| Bulgaria | 10 | 1 | 1 | 3.04 | Accept | 0.51 | 1 | 0.48 | 2 | N/A |
| Cyprus | 12 | 1 | 0 | 0.42 | Accept | 0.92 | 2 | 0.41 | 1 | N/A |
| Czech Rep. | 10 | 1 | 1 | 18.90 | Reject | 0.80 | 2 | 0.99 | 1 | 0.07 |
| Denmark | 22 | 1 | 1 | 2.52 | Accept | 0.75 | 2 | 0.88 | 3 | N/A |
| Estonia | 12 | 1 | 0 | 198.6 | Reject | 0.99 | 2 | 0.99 | 2 | N/A |
| Finland | 22 | 1 | 1 | 5.84 | Reject | 0.90 | 2 | 0.99 | 1 | <0.001 |
| France | 16 | 1 | 1 | 3.85 | Undecided | 0.84 | 1 | 0.99 | 2 | N/A |
| Germany | 22 | 1 | 0 | 0.82 | Accept | 0.79 | 1 | 0.64 | 1 | N/A |
| Greece | 22 | 1 | 0 | 4.51 | Undecided | 0.82 | 2 | 0.18 | 2 | N/A |
| Hungary | 9 | 1 | 0 | 340.0 | Reject | 0.98 | 1 | 0.81 | 2 | N/A |
| Italy | 22 | 1 | 1 | 4.72 | Undecided | 0.96 | 2 | 0.95 | 1 | 0.04 |
| Latvia | 9 | 0 | 1 | 2.52 | Accept | 0.73 | 1 | 0.84 | 2 | N/A |
| Lithuania | 13 | 0 | 1 | 4.46 | Undecided | 0.25 | 1 | 0.70 | 1 | N/A |
| Luxembourg | 22 | 1 | 1 | 5.10 | Reject | 0.90 | 1 | 0.17 | 1 | 0.02 |
| Malta | 10 | 1 | 0 | 6.52 | Reject | 0.91 | 1 | 0.95 | 2 | N/A |
| Netherlands | 22 | 1 | 1 | 3.73 | Accept | 0.55 | 1 | 0.97 | 3 | N/A |
| Poland | 16 | 1 | 0 | 77.1 | Reject | 0.93 | 3 | 0.91 | 4 | N/A |
| Portugal | 22 | 1 | 1 | 13.86 | Reject | 0.98 | 1 | 0.99 | 1 | 0.01 |
| Romania | 16 | 1 | 1 | 2.97 | Accept | 0.36 | 1 | 0.66 | 2 | N/A |
| Slovakia | 15 | 1 | 1 | 13.39 | Reject | 0.97 | 1 | 0.59 | 2 | 0.44 |
| Slovenia | 13 | 0 | 1 | 6.34 | Reject | 0.96 | 2 | 0.18 | 2 | N/A |
| Spain | 22 | 1 | 1 | 0.84 | Accept | 0.94 | 3 | 0.97 | 1 | N/A |
| Sweden | 22 | 1 | 1 | 1.87 | Accept | 0.77 | 1 | 0.95 | 4 | N/A |
| UK | 22 | 1 | 1 | 9.22 | Reject | 0.77 | 1 | 0.95 | 4 | 0.02 |
| Norway | 22 | 1 | 1 | 3.90 | Undecided | 0.71 | 1 | 0.65 | 1 | N/A |
| Switzerland | 21 | 1 | 1 | 7.77 | Reject | 0.92 | 2 | 0.95 | 3 | 0.09 |

**Appendix: Table A3b – Elliott Rothenburg Stock Unit root, Cointegration and Granger-causality test results**

| Country | N | ADF Unit Root Test Integration order (lag) $\alpha = 0.05$ | | Bounds Cointegration Test $H_0$: No Cointegration F statistic $\alpha = 0.05$ | | Causality Direction | | | | Error correction term (significance level) |
| | | | | | | Asylum -> GDP | | GDP -> Asylum | | |
| | | Asylum | GDP | Accept < 3.79 Reject > 4.85 | Test Result | F-prob. | Lag | F-prob. | Lag | |
|---|---|---|---|---|---|---|---|---|---|---|
| Ireland | 12 | 0 | 0 | N/A | - | 0.96 | 2 | 0.93 | 4 | N/A |
| Austria | 22 | 1 | 1 | 2.96 | Accept | 0.91 | 1 | 0.18 | 1 | N/A |
| Belgium | 22 | 0 | 0 | N/A | - | 074 | 1 | 0.90 | 2 | N/A |
| Bulgaria | 10 | 0 | 0 | N/A | - | 0.82 | 1 | 0.65 | 1 | N/A |
| Cyprus | 12 | 0 | 0 | N/A | - | 0.68 | 2 | 0.87 | 2 | N/A |
| Czech Rep. | 10 | 0 | 0 | N/A | - | 0.99 | 1 | 0.98 | 2 | 0.07 |
| Denmark | 22 | 1 | 1 | 2.52 | Accept | 0.75 | 2 | 0.88 | 3 | N/A |
| Estonia | 12 | 0 | 0 | N/A | - | 0.93 | 3 | 0.94 | 2 | N/A |
| Finland | 22 | 0 | 1 | 5.84 | Reject | 0.73 | 1 | 0.99 | 2 | 0.03 |
| France | 16 | 0 | 1 | 3.85 | Undecided | 0.84 | 1 | 0.99 | 2 | N/A |
| Germany | 22 | 0 | 0 | 0.82 | Accept | 0.24 | 2 | 0.87 | 1 | N/A |
| Greece | 22 | 0 | 0 | 4.51 | Undecided | 0.81 | 1 | 0.96 | 1 | N/A |
| Hungary | 9 | 1 | 1(2) | 340.0 | Reject | 0.99 | 2 | 0.89 | 1 | 0.03 |
| Italy | 22 | 0 | 0 | N/A | - | 0.91 | 3 | 0.89 | 2 | 0.04 |
| Latvia | 9 | 0 | 1 | 2.52 | Accept | 0.73 | 1 | 0.84 | 2 | N/A |
| Lithuania | 13 | 0 | 0 | N/A | - | 0.38 | 3 | 0.96 | 1 | N/A |
| Luxembourg | 22 | 0 | 1 | 5.10 | Reject | 0.90 | 1 | 0.41 | 1 | 0.02 |
| Malta | 10 | 0 | 1(2) | 6.52 | Reject | 1.0 | 1 | 0.99 | 1 | N/A |
| Netherlands | 22 | 0 | 1 | 3.73 | Accept | 0.62 | 1 | 0.97 | 4 | N/A |
| Poland | 16 | 0 | 0 | N/A | - | 0.99 | 1 | 0.96 | 1 | N/A |
| Portugal | 22 | 0 | 1 | 13.86 | Reject | 0.97 | 3 | 0.95 | 2 | 0.02 |
| Romania | 16 | 0 | 0 | N/A | - | 0.74 | 1 | 0.98 | 1 | N/A |
| Slovakia | 15 | 0 | 1 | 13.39 | Reject | 0.97 | 1 | 0.59 | 2 | 0.44 |
| Slovenia | 13 | 0 | 1 | 6.34 | Reject | 0.95 | 1 | 0.89 | 2 | N/A |
| Spain | 22 | 0 | 1 | 0.84 | Accept | 0.93 | 3 | 0.95 | 1 | N/A |
| Sweden | 22 | 0 | 0 | N/A | - | 0.95 | 3 | 0.52 | 2 | N/A |
| UK | 22 | 0 | 0 | N/A | - | 0.96 | 1 | 0.88 | 4 | 0.02 |
| Norway | 22 | 0 | 1 | 3.90 | Undecided | 0.74 | 4 | 0.24 | 1 | N/A |
| Switzerland | 21 | 0 | 1 | 7.77 | Reject | 0.95 | 3 | 0.52 | 2 | N/A |

**Appendix: Table A4 - Smoothed per capita Asylum applications to GDP elasticises in EEA countries relative to Ireland**

| Country | Year 1998 | 1999 | 2000 | 2001 | 2002 | 2003 | 2004 | 2005 | 2006 | 2007 | 2008 | 2009 | Mean |
|---|---|---|---|---|---|---|---|---|---|---|---|---|---|
| Ireland | 1 | 1 | 1 | 1 | 1 | 1 | 1 | 1 | 1 | 1 | 1 | 1 | 1 |
| Cyprus | | | | 0.86 | 0.77 | 1.08 | 1.44 | 1.48 | 1.40 | 1.47 | 1.37 | 1.29 | 1.17 |
| Sweden | 0.96 | 0.88 | 0.84 | 0.88 | 0.90 | 1.01 | 1.14 | 1.14 | 1.18 | 1.26 | 1.26 | 1.26 | 1.04 |
| Norway | 0.94 | 0.90 | 0.85 | 0.88 | 0.90 | 0.98 | 1.04 | 1.01 | 1.02 | 1.05 | 1.19 | 1.24 | 0.99 |
| Switzerland | 1.02 | 0.97 | 0.83 | 0.82 | 0.83 | 0.90 | 1.02 | 1.02 | 1.02 | 1.03 | 1.13 | 1.14 | 0.97 |
| Belgium | 0.98 | 0.97 | 0.92 | 0.86 | 0.81 | 0.84 | 0.95 | 1.00 | 0.97 | 1.00 | 1.07 | 1.16 | 0.95 |
| Malta | 0.00 | 0.00 | 0.62 | 0.60 | 0.67 | 0.79 | 1.04 | 1.13 | 1.18 | 1.20 | 1.34 | 1.35 | 0.94 |
| Austria | 0.87 | 0.86 | 0.78 | 0.84 | 0.88 | 0.95 | 1.06 | 1.09 | 1.02 | 0.99 | 1.01 | 1.06 | 0.94 |
| Denmark | 0.93 | 0.88 | 0.84 | 0.87 | 0.81 | 0.87 | 0.98 | 0.98 | 0.96 | 0.98 | 0.99 | 1.04 | 0.92 |
| UK | 0.90 | 0.88 | 0.83 | 0.82 | 0.84 | 0.88 | 0.93 | 0.91 | 0.90 | 0.91 | 0.93 | 0.94 | 0.88 |
| Netherlands | 1.02 | 0.92 | 0.85 | 0.81 | 0.73 | 0.75 | 0.81 | 0.87 | 0.93 | 0.83 | 0.94 | 0.98 | 0.87 |
| Slovakia | 0.61 | 0.68 | 0.67 | 0.91 | 0.96 | 1.07 | 1.26 | 1.05 | 1.00 | 0.94 | 0.64 | 0.57 | 0.86 |
| Finland | 0.69 | 0.75 | 0.71 | 0.62 | 0.70 | 0.76 | 0.91 | 0.97 | 0.92 | 0.81 | 0.99 | 1.11 | 0.81 |
| France | 0.73 | 0.70 | 0.66 | 0.67 | 0.67 | 0.75 | 0.89 | 0.91 | 0.90 | 0.94 | 0.93 | 0.94 | 0.79 |
| Germany | 1.00 | 0.91 | 0.81 | 0.80 | 0.76 | 0.75 | 0.72 | 0.68 | 0.64 | 0.60 | 0.65 | 0.72 | 0.76 |
| Poland | 0.67 | 0.58 | 0.58 | 0.59 | 0.60 | 0.70 | 0.85 | 0.85 | 0.82 | 0.91 | 1.00 | 1.11 | 0.75 |
| Greece | 0.59 | 0.41 | 0.44 | 0.53 | 0.55 | 0.66 | 0.64 | 0.81 | 0.92 | 1.07 | 1.07 | 1.02 | 0.70 |
| Czech Rep. | | | 0.72 | 0.82 | 0.72 | 0.82 | 0.77 | 0.71 | 0.67 | 0.49 | 0.48 | 0.42 | 0.66 |
| Hungary | | | | 0.80 | 0.72 | 0.56 | 0.47 | 0.50 | 0.60 | 0.70 | 0.72 | 0.81 | 0.63 |
| Luxembourg | 0.62 | 0.60 | 0.53 | 0.52 | 0.51 | 0.56 | 0.67 | 0.68 | 0.66 | 0.68 | 0.69 | 0.69 | 0.61 |
| Spain | 0.54 | 0.53 | 0.49 | 0.49 | 0.46 | 0.51 | 0.58 | 0.60 | 0.59 | 0.59 | 0.64 | 0.68 | 0.55 |
| Bulgaria | | | | 0.63 | 0.72 | 0.79 | 0.58 | 0.43 | 0.32 | 0.26 | 0.37 | 0.37 | 0.43 | 0.50 |
| Italy | 0.38 | 0.44 | 0.43 | 0.45 | 0.45 | 0.48 | 0.52 | 0.53 | 0.54 | 0.57 | 0.65 | 0.66 | 0.50 |
| Lithuania | | | 0.32 | 0.39 | 0.37 | 0.41 | 0.13 | -0.01 | 0.15 | 0.09 | 0.52 | 0.55 | 0.30 |
| Latvia | | | | -0.17 | -0.02 | -0.29 | -0.50 | -0.17 | -0.14 | 0.14 | 0.35 | 0.43 | 0.27 |
| Portugal | 0.44 | 0.34 | 0.26 | 0.25 | 0.26 | 0.16 | 0.16 | 0.19 | 0.25 | 0.32 | 0.29 | 0.26 | 0.27 |
| Slovenia | 0.24 | 0.23 | 0.23 | 0.23 | 0.22 | 0.25 | 0.30 | 0.31 | 0.30 | 0.30 | 0.30 | 0.29 | 0.26 |
| Romania | 0.42 | 0.41 | 0.37 | 0.44 | 0.19 | 0.13 | -0.09 | -0.07 | -0.06 | 0.09 | 0.33 | 0.26 | 0.22 |
| Estonia | | | | | 0.02 | 0.13 | 0.03 | 0.06 | -0.09 | 0.18 | 0.26 | 0.41 | 0.10 |